\def\webDOI{http://dx.doi.org}

\documentclass[final,2p,times]{amsart}

\usepackage{amssymb}
\usepackage{amsmath}
\usepackage{amsfonts}

\usepackage[pdftex,pdfborderstyle={/S/U/W 1}]{hyperref}
\usepackage{cite}
\usepackage{rotating}

\newcommand{\Rset}{\mathbb{R}}
\newcommand{\id}{\mathrm{d}}
\newcommand{\iexp}{\mathrm{e}}
\newcommand{\xj}{x}

\newcommand{\Poly}{P}
\newcommand{\Qoly}{Q}
\newcommand{\Roly}{R}
\newcommand{\Jacobi}{J}
\newcommand{\Jacab}[3]{\Jacobi_{#1}^{(#2,#3)}}
\newcommand{\Hermite}{H}
\newcommand{\Laguerre}{L}
\newcommand{\Gegen}[2]{C_{#1}^{(#2)}}

\newcommand{\coef}{b}
\newcommand{\Coef}{B}
\newcommand{\weight}{\mu}
\newcommand{\support}{I}
\newcommand{\scalar}[1]{\langle#1\rangle}
\newcommand{\norm}[1]{\hat{#1}}
\newcommand{\Polyn}{\norm{\Poly}}

\newcommand{\demi}{\frac{1}{2}}
\newcommand{\eref}[1]{Eq.~(\ref{#1})}
\newcommand{\sref}[1]{Sect.~\ref{#1}}
\newcommand{\Matlab}{\href{http://fr.mathworks.com/products/matlab/}{\texttt{Matlab}}}

\def\og{\leavevmode\raise.3ex\hbox{$\scriptscriptstyle\langle\!\langle$~}}
\def\fg{\leavevmode\raise.3ex\hbox{~$\!\scriptscriptstyle\,\rangle\!\rangle$}}

\begin{document}

\title[Higher-order moments of PC expansions]{Computation of Higher-Order Moments of Generalized Polynomial Chaos Expansions}

\author[\'E. Savin]{\'Eric Savin}
\address[\'E. Savin]{Onera--The French Aerospace Lab, F-92322 Ch\^atillon cedex, France}
\thanks{Corresponding author: \'E. Savin, Onera--The French Aerospace Lab, Computational Fluid Dynamics Dept., 29, avenue de la Division Leclerc, F-92322 Ch\^atillon cedex, France (Eric.Savin@onera.fr).}
\email{eric.savin@onera.fr}

\author[B. Faverjon]{B\'eatrice Faverjon}
\address[B. Faverjon]{INSA-Lyon, F-69621 Villeurbanne cedex, France}
\email{beatrice.faverjon@insa-lyon.fr}

\begin{abstract}
Because of the complexity of fluid flow solvers, non-intrusive uncertainty quantification techniques have been developed in aerodynamic simulations in order to compute the quantities of interest required in an optimization process, for example. The objective function is commonly expressed in terms of moments of these quantities, such as the mean, standard deviation, or even higher-order moments. Polynomial surrogate models based on polynomial chaos expansions have often been implemented in this respect. The original approach of uncertainty quantification using polynomial chaos is however intrusive. It is based on a Galerkin-type formulation of the model equations to derive the governing equations for the polynomial expansion coefficients. Third-order, indeed fourth-order moments of the polynomials are needed in this analysis. Besides, both intrusive and non-intrusive approaches call for their computation provided that higher-order moments of the quantities of interest need be post-processed. In most applications they are evaluated by Gauss quadratures, and eventually stored for use throughout the computations. In this paper analytical formulas are rather considered for the moments of the continuous polynomials of the Askey scheme, so that they can be evaluated by quadrature-free procedures instead. \Matlab\ codes have been developed for this purpose and tested by comparisons with Gauss quadratures.

\end{abstract}

\keywords{Orthogonal polynomials, Linearization problem, Polynomial chaos, Uncertainty quantification}

\maketitle

\section{Introduction}\label{sec:intro}

The polynomial chaos (PC), or homogeneous chaos expansion defined as the span of Hermite polynomial functionals of a Gaussian random variable has been introduced by Wiener \cite{WIE38} for stochastic processes. Mean-square convergence is guaranteed by the Cameron-Martin theorem \cite{CAM47} and is optimal (\emph{i.e}. exponential) for Gaussian processes. For arbitrary random processes the numerical study in \cite{XIU02} has shown that the convergence rates are not optimal. This observation has prompted the development of generalized chaos expansions (gPC) involving other families of polynomials \cite{XIU02,SOI04}. They consist in expanding any function of random variables into a linear combination of orthogonal polynomials with respect to the probability density functions of these underlying random variables. The homogeneous and generalized homogeneous chaos expansions have recently received a broad attention in engineering sciences, where they are extensively used as a constructive tool for representing random vectors, matrices, tensors or fields for the purpose of quantifying uncertainty in complex systems. Several applications are described in \emph{e.g.} \cite{CLO01,CRE06,DEB05,GHA99,GHA03,GHI02,NAJ09,OLM10,SAV16,SEG13,SOI04,SUN79,XIU02,XIU05} and references therein.

Complex aerodynamic analysis and design of aircraft use high-fidelity computational fluid dynamics (CFD) tools for shape optimization for example, whereby some robustness is achieved by considering uncertain operational, environmental, or manufacturing parameters. Non-intrusive uncertainty propagation is typically considered in CFD, because the complex flow solvers are preferably treated as black boxes in order to compute the output quantities of interest that are required to evaluate the objective function of an optimization process. The latter is often expressed in terms of moments of the quantities of interest, such as the mean, standard deviation, or even higher-order moments (skewness, kurtosis...). Together with the Monte-Carlo method, the method of moments, the stochastic collocation approach or polynomial chaos expansions are widely used non-intrusive approaches for evaluating stochastic objective functions. Using the latter approach, the computation of higher-order moments of the output quantities of interest in an optional post-processing step requires the evaluation of higher-order moments of the orthogonal polynomials involved in these expansions.

The intrusive approach originally introduced in \cite{GHA03,SUN79} is based on a Galerkin-type projection formulation of the model equations, typically partial differential equations, to derive the governing equations for the spectral expansion coefficients of the output quantities of interest. More precisely, the PC expansions of the model parameters and variables are substituted in the model equations, which in turn yield the evolution equations for the outputs from Galerkin projections using the orthogonal polynomials of the PC expansions. This procedure gives rise to third-order, even fourth-order moments of these polynomials as illustrated with some simple examples in \cite{DEB05,GHA03,NAJ09}. Their computation is needed at this stage, while they may also be useful in a post-processing step if higher-order moments of the output quantities of interest are requested. The polynomial moments are evaluated by Gauss quadratures in most applications, and then stored for use throughout the computations.

The main purpose of this communication is to show that analytical formulas for these moments are available, so that they could be evaluated numerically by general, quadrature-free procedures instead. A numerical implementation of these formulas is thus proposed in the form of freely available \Matlab\ codes. It is believed that such results may have some relevance for the engineering community interested in uncertainty quantification issues, using either intrusive or non-intrusive approaches based on PC and gPC expansions. The use of higher-order moments in post-processing steps for example is illustrated in \cite{SIN10} for optimization, or in \cite{GER16} for global sensitivity analysis, among other possible applications. The paper is organized as follows. The standard linearization problem of a product of polynomials is very briefly introduced in the next section. Then it is applied to the computation of higher-order moments of orthogonal polynomials in \sref{sec:moments}, where the available explicit expressions of the third-order moments are listed for Jacobi, generalized Hermite, and generalized Laguerre polynomials (thus covering all continuous polynomials identified in \cite{CHI78,XIU02} for example). Numerical implementation of these results using \Matlab\ is addressed in \sref{sec:num}. It is validated by comparisons with classical evaluations of the third-order moments by Gauss quadratures, for which the codes used in this process are also provided. Some conclusions and perspectives are finally drawn in \sref{sec:CL}.

\section{Standard linearization problem}\label{sec:slp}

Let $\Qoly_j(\xj)$ and $\Roly_k(\xj)$ be two polynomials of degrees $j$ and $k$ respectively. Let $\smash{\{\Poly_n\}_{n\geq 0}}$ be an arbitrary sequence of polynomials such that $\deg\Poly_n=n$. The general linearization problem consists in finding the coefficients $\Coef_n(j,k)$ such that:
\begin{equation}\label{eq:glp}
\Qoly_j(\xj)\Roly_k(\xj)=\sum_{n=0}^{j+k}\Coef_n(j,k)\Poly_n(\xj)\,.
\end{equation}
A particular case of this problem is the standard linearization problem (or Clebsch-Gordan-type problem) for which $\Qoly_n\equiv\Roly_n\equiv\Poly_n$:
\begin{equation}\label{eq:slp}
\Poly_j(\xj)\Poly_k(\xj)=\sum_{n=0}^{j+k}\Coef_n(j,k)\Poly_n(\xj)\,.
\end{equation}
Another particular case is the so-called connection problem, for which $\Roly_k(\xj)=1$; if in addition $\Qoly_j(\xj)=\xj^j$ is chosen, it is referred to as the inversion problem for the sequence $\smash{\{\Poly_n\}_{n\geq 0}}$. These problems have been the subject of numerous investigations, some of them being addressed in \cite{ABD15,ADA78,BUS39,CHA07,CHA10,CHA11,FEL38,GAS70,HYL62,RAH81,SAN99,WAT38} and references therein. The objective of this communication is definitely not to review exhaustively these results, but to apply them to the computation of higher-order moments from PC or gPC expansions of random parameters and/or functionals. Such expansions have emerged as efficient numerical tools for uncertainty quantification and propagation in complex engineering systems. They have been considered in the intrusive spectral stochastic finite element method proposed in \cite{GHA03,SUN79}, or in the non-intrusive stochastic collocation method proposed in \cite{GHI02,XIU05}, among other possible applications.

\section{Higher-order moments of orthonormal polynomials}\label{sec:moments}

Consider now the standard linearization problem for the family $\smash{\{\Poly_n\}_{n\geq 0}}$ of orthogonal polynomials with respect to the non-negative density $\xj\mapsto\weight(\xj)$ of support $\support$, \emph{i.e.}:
\begin{equation}\label{eq:orthonorm}
\scalar{\Poly_m\Poly_n}_\weight:=\int_\support\Poly_m(\xj)\Poly_n(\xj)\weight(\xj)\id\xj=\gamma_n\delta_{mn}\,,\quad m,n\geq 0\,,
\end{equation}
where $\smash{\delta_{mn}}$ is the usual Kronecker symbol, and $\smash{\gamma_n}>0$ is the normalization constant. Then clearly from \eref{eq:slp} the following holds:
\begin{equation}\label{eq:m3}
\scalar{\Polyn_j\Polyn_k\Polyn_l}_\weight=\sqrt{\frac{\gamma_l}{\gamma_j\gamma_k}}\Coef_l(j,k):=\coef_l(j,k)\,,\quad l\leq j+k\,,
\end{equation}
if we introduce the orthonormalized polynomials $\smash{\Polyn_n\equiv\gamma_n^{-\demi}\Poly_n}$. The roles of $j,k$ and $l$ in \eref{eq:m3} are transparent so they can be permutated in this formula. This should be apparent in the analytical expression of $\coef_l(j,k)$ whenever it is available. In addition, one has $\coef_l(j,k)=0$ whenever $l<|j-k|$. Indeed, either $k+l<j$ or $j+l<k$ in this case thus $\smash{\deg\{\Poly_k\Poly_l\}}<\smash{\deg\{\Poly_j\}}$ or $\smash{\deg\{\Poly_j\Poly_l\}}<\smash{\deg\{\Poly_k\}}$, and consequently $\smash{\scalar{\Poly_j\Poly_k\Poly_l}_\weight}=0$. The fourth-order moment can be derived from the above third-order moments by simple mathematical induction:
\begin{equation}
\begin{split}
\scalar{\Polyn_j\Polyn_k\Polyn_l\Polyn_m}_\weight &=\sum_{n=0}^{j+k}\coef_n(j,k)\scalar{\Polyn_l\Polyn_m\Polyn_n}_\weight\\
&=\sum_{n=0}^{j+k}\coef_n(j,k)\coef_n(l,m)\,.
\end{split}
\end{equation}
Likewise, free permutations of the transparent indices $j,k,l$ and $m$ are applicable. Higher-order moments are obtained along the same lines by repeated uses of \eref{eq:slp} and induction. The above third-order and fourth-order moments of orthonormal polynomials typically arise in the determination of the PC expansion for the product of two or three stochastic variables, as illustrated in \cite{DEB05} for example. Here the third order tensor $C_{jkl}=\scalar{\Poly_j\Poly_k\Poly_l}_\weight$ is rather evaluated numerically by dedicated quadrature rules, benefiting to some extent from its sparsity. As explained in the introductory section, these moments are needed in the spectral stochastic finite element method for example: the PC expansions for model parameters and variables are substituted into the governing equations, then using a Galerkin projection method evolution equations are obtained for the spectral coefficients in the PC expansions. In non-intrusive stochastic collocation methods, the spectral coefficients of the PC expansions of the output quantities of interest are computed by running the underlying physical model for particular model parameter values, typically belonging to an adapted quadrature set \cite{SAV16}. Both in the intrusive and non-intrusive methods, the moments are used to post-process the PC expansions for deriving the moments of the output quantities of interest, so long as they are needed.

The linearization coefficients for some classical families of orthogonal polynomials are explicitly given in the subsequent sections. Jacobi, generalized Hermite, and generalized Laguerre polynomials are more particularly addressed. Families corresponding to discrete non-negative measures $\weight(\id\xj)$ may be considered alike, though they are not reviewed in this communication.

\subsection{Jacobi polynomials}

The Jacobi polynomials $\smash{\{\Jacab{n}{\alpha}{\beta}\}_{n\geq 0}}$ are orthogonal with respect to the weight function $\weight(\xj)=\smash{(1-\xj)^\alpha(1+\xj)^\beta}$, with $\alpha,\beta>-1$ and $I=[-1,1]$. They are defined by \emph{e.g.} the standard Rodrigues' formula:
\begin{displaymath}
\begin{split}
\Jacab{n}{\alpha}{\beta}(\xj) &=\frac{1}{\mu(\xj)}\frac{\id^n}{\id\xj^n}\left(\frac{\mu(\xj)}{n!}\left(\frac{\xj^2-1}{2}\right)^n\right) \\
&=\sum_{j=0}^n\binom{n+\alpha}{n-j}\binom{n+\beta}{j}\left(\frac{\xj-1}{2}\right)^{j}\left(\frac{\xj+1}{2}\right)^{n-j}\,,
\end{split}
\end{displaymath}
where:
\begin{displaymath}
\binom{z}{p}=\frac{z!}{(z-p)!p!}
\end{displaymath}
stands for the generalized binomial coefficient. Indeed, one has $z!:=\Gamma(z+1)=\smash{\int_0^{+\infty}t^z\iexp^{-t}\id t}$ and $\Gamma(p+1)=p!$, the usual factorial, if $p$ is an integer. Jacobi polynomials arise in gPC expansions for random variables following beta distributions of the first kind; see \emph{e.g.} \cite{SAV16,XIU02}. The normalization constant $\gamma_n$ in \eref{eq:orthonorm} then reads:
\begin{displaymath}
\gamma_n=\frac{2^{\alpha+\beta+1}(n+\alpha)!(n+\beta)!}{(2n+\alpha+\beta+1)(n+\alpha+\beta)!n!}\,.
\end{displaymath}
The linearization coefficients $\Coef_n(j,k)$ in the general linearization problem:
\begin{displaymath}
\Jacab{j}{\lambda}{\delta}(\xj)\Jacab{k}{\mu}{\nu}(\xj)=\sum_{n=|j-k|}^{j+k}\Coef_n(j,k)\Jacab{n}{\alpha}{\beta}(\xj)\,,\quad\alpha,\beta,\lambda,\delta,\mu,\nu>-1\,,
\end{displaymath}
are given in \cite[Eq.~(12)]{CHA10} in terms of double hypergeometric functions (the so-called Kamp\'e de F\'eriet functions). In the context of PC expansions we are rather interested in the standard linearization problem for which $\alpha=\lambda=\mu$ and $\beta=\delta=\nu$. A representation in terms of generalized hypergeometric series was derived in \cite{ABD15,RAH81} for this problem. For numerical robustness, we will rather resort to the older induction formula derived in~\cite{GAS70}. Here the linearization coefficients are given by:
\begin{equation}\label{eq:LinCoef_Jacobi}
\Coef_n(j,k)=(-1)^{j+k+n}\frac{k!}{(k+\beta)!}\tilde{\Coef}_n(j,k)\,,
\end{equation}
where the coefficients $\tilde{\Coef}_n(j,k)$ are obtained by the induction formula~\cite[Eq.~(4.13)]{HYL62}:
\begin{multline*}
\frac{[(j+k+\alpha+\beta+1)^2-(n+\alpha+\beta)^2][(n+\alpha+\beta)^2-(j-k)^2]}{(2n+\alpha+\beta)[2(n-1)+\alpha+\beta+1]}(n+\beta)\tilde{\Coef}_{n-1}(j,k) \\
-\frac{[(j+k+\alpha+\beta+1)^2-(n+1)^2][(n+1)^2-(j-k)^2]}{[2(n+1)+\alpha+\beta][2(n+1)+\alpha+\beta+1]}(n+1+\alpha)\tilde{\Coef}_{n+1}(j,k) \\
+\frac{[(j+k+\alpha+\beta+1)^2-(n+1+\alpha+\beta)^2][(n+1)^2-(j-k)^2]}{2(n+1)+\alpha+\beta}\left(\frac{\beta-\alpha}{2}\right)\tilde{\Coef}_n(j,k) \\
-\frac{[(j+k+\alpha+\beta+1)^2-(n+\alpha+\beta)^2][n^2-(j-k)^2]}{2n+\alpha+\beta}\left(\frac{\beta-\alpha}{2}\right)\tilde{\Coef}_n(j,k) =0\,,
\end{multline*}
starting from (assuming $j\geq k$):
\begin{displaymath}
\begin{split}
\tilde{\Coef}_{j-k-1}(j,k) &=0\,,\\
\tilde{\Coef}_{j-k}(j,k) &=\frac{[2(j-k)+\alpha+\beta+1]!(2k+\alpha+\beta)!(j+\alpha)!(j+\beta)!}{(2j+\alpha+\beta+1)!(k+\alpha+\beta)!(j-k+\alpha)!(j-k)!j!}\,.
\end{split}
\end{displaymath}

The ultraspherical (Gegenbauer) polynomials $\smash{\{\Gegen{n}{\lambda}\}_{n\geq 0}}$ correspond to the particular case $\alpha=\beta=\smash{\lambda-\demi}$ with $\lambda\neq 0$ and the standardization:
\begin{displaymath}
\Gegen{n}{\lambda}(\xj)=\frac{(2\lambda)_n}{(\lambda+\demi)_n}\Jacab{n}{\lambda-\demi}{\lambda-\demi}(\xj)\,,\quad\lambda>-\demi\,,
\end{displaymath}
where $\smash{(z)_n}:=\smash{\frac{\Gamma(z+n)}{\Gamma(z)}}$ stands for the usual Pochhammer symbol. The corresponding linearization coefficients $\Coef_n(j,k)$ such that:
\begin{displaymath}
\Gegen{j}{\lambda}(\xj)\Gegen{k}{\lambda}(\xj)=\sum_{n=0}^{\min(j,k)}\Coef_{j+k-2n}(j,k)\Gegen{j+k-2n}{\lambda}(\xj)\,,\quad\lambda>-\demi\,,
\end{displaymath}
are given by the Dougall's formula \cite[Eq.~(5.7)]{ASK75} (see also \cite[Eq.~(28)]{CHA10}):
\begin{multline}\label{eq:Dougall}
\Coef_{j+k-2n}(j,k)=\frac{(\lambda+j+k-2n)(j+k-2n)!}{(\lambda+j+k-n)n!(j-n)!(k-n)!} \\
\times\frac{(2\lambda)_{j+k-n}(\lambda)_{j-n}(\lambda)_{k-n}(\lambda)_n}{(2\lambda)_{j+k-2n}(\lambda)_{j+k-n}}\,.
\end{multline}
For the family of Legendre polynomials such that $\smash{\lambda=\demi}$ the Neumann-Adams formula \cite[p. 91]{ADA78,NEU78} is recovered, namely:
\begin{equation}\label{eq:Adams}
\Coef_{j+k-2n}(j,k)=\frac{2(j+k-2n)+1}{2(j+k-n)+1}\frac{(j+k-n)!(\demi)_{j-n}(\demi)_{k-n}(\demi)_n}{(\demi)_{j+k-n}(j-n)!(k-n)!n!}\,,
\end{equation}
where $\smash{(\demi)_n}=\smash{\frac{(2n)!}{4^nn!}}$, \emph{etc}. Legendre polynomials arise in gPC expansions for the important case of uniform distributions.

Lastly, Chebyshev polynomials of the first kind $\smash{\{T_n\}_{n\geq 0}}$ correspond to the special case $\alpha=\beta=-\demi$ and are:
\begin{displaymath}
T_n(\xj)=\cos(n\arccos\xj)=\frac{\Jacab{n}{-\demi}{-\demi}(\xj)}{\Jacab{n}{-\demi}{-\demi}(1)}\,,\quad\Jacab{n}{-\demi}{-\demi}(1)=\frac{1}{n!}\left(\demi\right)_n\,.
\end{displaymath}
Since:
\begin{displaymath}
2T_j(\xj)T_k(\xj)=T_{|j-k|}(\xj)+T_{j+k}(\xj)\,,
\end{displaymath}
the linearization coefficients are simply $\Coef_{|j-k|}(j,k)=\Coef_{j+k}(j,k)=\demi$ and $\Coef_n(j,k)=0$ otherwise.

\subsection{Hermite polynomials}

The generalized Hermite polynomials $\smash{\{\Hermite_n^{(\alpha)}\}_{n\geq 0}}$ are orthogonal with respect to the weight function $\mu(|\xj|)$ with $\weight(\xj)=\smash{\xj^{2\alpha}\iexp^{-\xj^2}}$, $\smash{\alpha>-\demi}$, on $I=\Rset$. They are given by the Rodrigues-like formula~\cite[p.~157]{CHI78}:
\begin{equation}\label{eq:RH}
\Hermite_n^{(\alpha)}(\xj)=\frac{1}{\mu(\xj)}\frac{\id^n}{\id\xj^n}\left(\mu(\xj)\xj^n K_n^{(\alpha)}(\xj)\right)\,,
\end{equation}
where:
\begin{displaymath}
\begin{split}
K_{2m}^{(\alpha)}(\xj) &=\frac{(-1)^m}{(\alpha+1)_m}\,{_1F_1}\left(\left.\begin{matrix}m \\ m+\alpha+1\end{matrix}\right|\xj^2\right)\,, \\
K_{2m+1}^{(\alpha)}(\xj) &=\frac{(-1)^m}{(\alpha+1)_{m+1}}\xj\,{_1F_1}\left(\left.\begin{matrix}m+1 \\ m+\alpha+2\end{matrix}\right|\xj^2\right)\,,
\end{split}
\end{displaymath}
and $\smash{_pF_q}$ is the generalized hypergeometric function defined as:
\begin{displaymath}
_pF_q\left(\left.\begin{matrix}({\boldsymbol a}_p) \\ ({\boldsymbol b}_q)\end{matrix}\right|\xj\right)=\sum_{k=0}^{\infty}\frac{(a_1)_k(a_2)_k\cdots(a_p)_k}{(b_1)_k(b_2)_k\cdots(b_q)_k}\frac{\xj^k}{k!}\,.
\end{displaymath}
The normalization constant $\gamma_n$ in \eref{eq:orthonorm} reads~\cite[p.~157]{CHI78}:
\begin{displaymath}
\gamma_n=2^{2n}\left\lfloor\frac{n}{2}\right\rfloor!\left(\left\lfloor\frac{n+1}{2}\right\rfloor+\alpha-\demi\right)!\,,
\end{displaymath}
where $\lfloor\cdot\rfloor$ is the largest integer function. This family reduces to the classical Hermite polynomials $\smash{\{\Hermite_n\}_{n\geq 0}}$ for $\alpha=0$. The latter arise in PC expansions for random variables following Gaussian distributions and are the original polynomial chaoses of the stochastic finite element method introduced in \cite{GHA03,SUN79}.  Rodrigues' formula (\ref{eq:RH}) for $\alpha=0$ and $\mu(\xj)=\smash{\iexp^{-\xj^2}}$ reads:
\begin{displaymath}
\Hermite_n(\xj)=\frac{1}{\mu(\xj)}\frac{\id^n}{\id\xj^n}\left((-1)^n\mu(\xj)\right)=n!\sum_{j=0}^{\lfloor\frac{n}{2}\rfloor}\frac{(-1)^j(2\xj)^{n-2j}}{(n-2j)!j!}\,,
\end{displaymath}
and the normalization constant is (owing to $\smash{(-\demi)}!=\sqrt{\pi}$):
\begin{displaymath}
\gamma_n=\sqrt{\pi}2^nn!\,.
\end{displaymath}
The linearization coefficients $\Coef_n(j,k)$ in the general linearization problem:
\begin{displaymath}
\Hermite_j^{(\lambda)}(\xj)\Hermite_k^{(\mu)}(\xj)=\sum_{n=0}^{\min(j,k)}\Coef_{j+k-2n}(j,k)\Hermite_{j+k-2n}^{(\alpha)}(\xj)\,,\quad\alpha,\lambda,\mu>-\demi\,,
\end{displaymath}
are given in \cite[Eq.~(3.23)]{CHA07} and \cite[Eq.~(3.5)]{CHA11} for the standardization of generalized Hermite polynomials $\smash{\{\mathcal{H}_n^{(\alpha)}\}_{n\geq 0}}$ introduced by Rosenblum \cite{ROS94}:
\begin{displaymath}
\mathcal{H}_n^{(\alpha)}(\xj)=\frac{n!}{\gamma_\alpha(n)}\Hermite_n^{(\alpha)}(\xj)\,,
\end{displaymath}
where $\smash{n\mapsto\gamma_\alpha(n)}$ plays the role of a generalized factorial:
\begin{displaymath}
\begin{split}
\gamma_\alpha(2m) &=2^{2m}m!\left(\alpha+\demi\right)_m\,, \\
\gamma_\alpha(2m+1) &=2^{2m+1}m!\left(\alpha+\demi\right)_{m+1}\,.
\end{split}
\end{displaymath}
Again, in the context of PC expansions we are rather interested in the standard linearization problem $\lambda=\mu=\alpha$, for which the linearization coefficients for the chosen standardization (\ref{eq:RH}) read: 
\begin{equation}\label{eq:LinCoef_Hermite}
\Coef_{j+k-2n}(j,k)=\frac{\gamma_\alpha(j)\gamma_\alpha(k)}{\gamma_\alpha(j+k-2n)}\sum_{p=0}^{\lfloor\frac{j}{2}\rfloor}\sum_{q=0}^{\lfloor\frac{k}{2}\rfloor}\frac{\gamma_\alpha(j+k-2(p+q))}{\gamma_\alpha(j-2p)\gamma_\alpha(k-2q)p!q!}\frac{(-n)_{p+q}}{n!}\,.
\end{equation}

The explicit linearization formula for classical Hermite polynomials $\alpha=0$ is known as the Feldheim's formula and reads \cite{FEL38} (see also \cite[Eq.~(3.10)]{CHA11}):
\begin{equation}
\Hermite_j(\xj)\Hermite_k(\xj)=\sum_{n=0}^{\min(j,k)}\binom{j}{n}\binom{k}{n} 2^n n!\Hermite_{j+k-2n}(\xj)\,,
\end{equation}
where $\binom{j}{n}=\smash{\frac{j!}{(j-n)!n!}}$ is the usual binomial coefficient for two integers $j\geq n$. We arrive at:
\begin{displaymath}
\Coef_l(j,k)=\frac{\sqrt{\pi}2^n\,j!k!l!}{(n-j)!(n-k)!(n-l)!}
\end{displaymath}
whenever $2n=j+k+l$ is even, and $l\leq j+k$, $k\leq j+l$, $j\leq k+l$; and $\Coef_l(j,k)=0$ otherwise. This formula agrees with \emph{e.g.} \cite[Eq.~(8)]{BUS39} or \cite[p.~273]{SUL16} up to a proper normalization of the Hermite polynomials.

\subsection{Laguerre polynomials}

The generalized Laguerre polynomials $\smash{\{\Laguerre_n^{(\alpha)}\}_{n\geq 0}}$ are orthogonal with respect to the weight function $\weight(\xj)=\smash{\xj^\alpha\iexp^{-\xj}}$, with $\smash{\alpha>-1}$ and $I=[0,+\infty($. They are defined by \emph{e.g.} the Rodrigues' formula:
\begin{displaymath}
\Laguerre_n^{(\alpha)}(\xj)=\frac{1}{\mu(\xj)}\frac{\id^n}{\id\xj^n}\left(\mu(\xj)\frac{\xj^n}{n!}\right)=\sum_{j=0}^n\binom{n+\alpha}{n-j}\frac{(-\xj)^j}{j!}\,,
\end{displaymath}
and arise in gPC expansions for random variables following gamma distributions; see \emph{e.g.} \cite{SEG13,XIU02}. This family reduces to the classical Laguerre polynomials $\smash{\{\Laguerre_n\}_{n\geq 0}}$ for $\alpha=0$, applicable to exponentially distributed random variables. The normalization constant $\gamma_n$ in \eref{eq:orthonorm} reads:
\begin{displaymath}
\gamma_n=\frac{(n+\alpha)!}{n!}\,. 
\end{displaymath}
The linearization coefficients $\Coef_n(j,k)$ in the general linearization problem:
\begin{displaymath}
\Laguerre_j^{(\lambda)}(\xj)\Laguerre_k^{(\mu)}(\xj)=\sum_{n=|j-k|}^{j+k}\Coef_{n}(j,k)\Laguerre_{n}^{(\alpha)}(\xj)\,,\quad\alpha,\lambda,\mu>-1\,,
\end{displaymath}
are given in \cite[Eq.~(3.24)]{CHA07} in terms of double hypergeometric functions. Again, in the context of PC expansions we are rather interested in the standard linearization problem $\lambda=\mu=\alpha$, for which the linearization coefficients are given by \cite{SAN99,WAT38} in terms of a terminating hypergeometric series $\smash{_3F_2}$:
\begin{displaymath}
\Coef_{j+k-n}(j,k)=\frac{(-2)^n}{n!}\frac{(j+k-n)!}{(j-n)!(k-n)!}{_3F_2}\left(\left.\begin{matrix}-\frac{n}{2},\,-\frac{n-1}{2},\,j+k-n+\alpha+1 \\ j-n+1,\,k-n+1\end{matrix}\right|1\right)\,.
\end{displaymath}
The first $\max(n-j,n-k)$ terms of the series above are ignored whenever $n>\max(j,k)$; thus:
\begin{multline}\label{eq:LinCoef_Laguerre}
\Coef_{j+k-n}(j,k)=\frac{(-2)^n(j+k-n)!}{(j+k-n+\alpha)!n!} \\
\times\sum_{l=\max(0,n-j,n-k)}^{\lfloor\frac{n}{2}\rfloor}\left(-\frac{n}{2}\right)_l\left(-\frac{n-1}{2}\right)_l\frac{(j+k-n+\alpha+l)!}{(j-n+l)!(k-n+l)!}\,.
\end{multline}

\section{Numerical implementation}\label{sec:num}

The various formulas above have been implemented in \Matlab. The routines are distributed under CeCILL-C license and are freely available at:
$$\text{\href{https://github.com/ericsavin/LinCoef/}{\texttt{https://github.com/ericsavin/LinCoef/}}.}$$
They were compared with the results obtained with classical Gauss quadratures for the computation of the third-order moments of \eref{eq:m3}. The Golub-Welsch algorithm \cite{GOL69} is used for computing Gauss quadrature weights and nodes. The recurrence coefficients for monic Jacobi, generalized Hermite, and generalized Laguerre polynomials in:
\begin{displaymath}
\Poly_{n+1}(\xj)=(\xj-c_n)\Poly_n(\xj)-d_n\Poly_{n-1}(\xj)
\end{displaymath}
are given in the table \ref{tb:recurrence} below, together with the leading-order coefficient $p_n$ and the zero-th moment $\smash{\weight_0}=\smash{\scalar{1}_\weight}$ for completeness. Computations by the analytical formulas detailed in the foregoing section are in very good agreement with Gauss quadratures, which validate our proposed codes.

\begin{sidewaystable}[h!]
\vskip350pt
\begin{center}
\begin{tabular}{|c||c|c|c|c|}
\hline
& \makebox[11em]{$c_n$} & \makebox[11em]{$d_n$} & \makebox[11em]{$p_n$} & \makebox[11em]{$\weight_0$} \\
\hline\hline
$\Jacab{n}{\alpha}{\beta}(\xj)$ & $\frac{\beta^2-\alpha^2}{(2n+\alpha+\beta)(2n+\alpha+\beta+2)}$ & $\frac{4n(n+\alpha)(n+\beta)(n+\alpha+\beta)}{(2n+\alpha+\beta)^2[(2n+\alpha+\beta)^2-1]}$ & $\frac{1}{2^n}\binom{2n+\alpha+\beta}{n}$ & $2^{\alpha+\beta+1}\frac{\alpha!\beta!}{(\alpha+\beta+1)!}$ \\
$\Hermite_n^{(\alpha)}(\xj)$ & $0$ & $\demi[n+\alpha(1-(-1)^n)]$ & $2^n$ & $(\alpha-\demi)!$ \\
$\Laguerre_n^{(\alpha)}(\xj)$ & $2n+\alpha+1$ & $n(n+\alpha)$ & $\frac{(-1)^n}{n!}$ & $\alpha!$ \\
\hline
\end{tabular}
\vskip10pt
\caption{Recurrence coefficients for monic Jacobi, generalized Hermite, and generalized Laguerre polynomials.}\label{tb:recurrence}
\end{center}
\end{sidewaystable}

The main function is \texttt{LinCoef.m} which computes the linearization coefficients for Jacobi, Gegenbauer, generalized Hermite, and generalized Laguerre polynomials of arbitrary parameters $\alpha$ and $\beta$. Chebyshev polynomials (Jacobi polynomials with $\alpha=\beta=-\smash{\demi}$) are also specifically addressed. Three routines are provided to compare the implementation with Gauss quadratures: \texttt{TestHermite.m}, \texttt{TestJacobi.m}, and \texttt{TestLaguerre.m}. These quadrature sets are constructed with the \texttt{GNodeWt.m} function, while the polynomials are evaluated at the quadrature nodes by the dedicated functions \texttt{PGHern.m}, \texttt{PJacn.m}, and \texttt{PGLagn.m}. 


%
%
%
%

\section{Conclusions}\label{sec:CL}

In this paper, we have presented the existing results for the computation of the so-called linearization coefficients for products of orthogonal polynomials of the Jacobi, generalized Hermite, and generalized Laguerre families. These coefficients correspond to the third-order moments of orthogonal polynomials, but they also serve for the computation of higher-order moments by induction. Therefore, they can be used in the intrusive and non-intrusive polynomial chaos expansion methods for uncertainty quantification of engineering systems, among other possible applications. In the intrusive approach the third-order or fourth-order moments arise from Galerkin-type projections of the governing equations of the system models and are needed to carry out the overall UQ analysis. In both intrusive and non-intrusive approaches, these moments pertain to the computation of higher-order moments (skewness, kurtosis and beyond) of the output quantities of interest in a post-processing step, so long as they are needed. These results have been implemented in \Matlab\ and the codes have been validated by comparison with usual Gauss quadratures. The present overview concerns continuous polynomials, but it can be extended to discrete polynomials alike.

\section*{Acknowledgements}
The work of \'E.S. has been partially supported by the European Union's Seventh Framework Programme for research, technological development and demonstration under grant agreement \#ACP3-GA-2013-605036 (UMRIDA Project \href{http://www.umrida.eu}{\texttt{www.umrida.eu}}).

\end{document}